\documentclass[pre,citeautoscript,superscriptaddress,twocolumn]{revtex4}

\usepackage{amsfonts,amsmath,amssymb}
\usepackage[dvips]{graphicx,color}

\begin{document}

\title{Rupture of a Biomembrane under Dynamic Surface Tension}

\author{D. J. Bicout}
\affiliation{Institut Laue-Langevin, 6 rue Jules Horowitz, B.P.
156, 38042 Grenoble, France} \affiliation{Biomathematics and
Epidemiology, EPSP - TIMC, UMR 5525 CNRS, Joseph Fourier
University, VetAgro Sup Lyon, 69280 Marcy l'Etoile, France}

\author{E. Kats}
\affiliation{Institut Laue-Langevin, 6 rue Jules Horowitz, B.P.
156, 38042 Grenoble, France} \affiliation{L. D. Landau Institute
for Theoretical Physics, RAS, 117940 GSP-1, Moscow, Russia}

\begin{abstract}
How long a fluid membrane vesicle stressed with a steady ramp of
micropipette last before rupture? Or conversely, how high the
surface tension should be to rupture a membrane? To answer these
challenging questions we have developed a theoretical framework
that allows description and reproduction of Dynamic Tension
Spectroscopy (DTS) observations. The kinetics of the membrane
rupture under ramps of surface tension is described as a
combination of initial pore formation followed by Brownian process
of the pore radius crossing the time-dependent energy barrier. We
present the formalism and derive (formal) analytical expression of
the survival probability describing the fate of the membrane under
DTS conditions. Using numerical simulations for the membrane
prepared in an initial state with a given distribution of times
for pore nucleation, we have studied the membrane lifetime (or
inverse of rupture rate) and distribution of membrane surface
tension at rupture as a function of membrane characteristics like
pore nucleation rate, the energy barrier to failure and tension
loading rate. It is found that simulations reproduce main features
of the experimental data, particularly, the pore nucleation and
pore size diffusion controlled limits of membrane rupture
dynamics. This approach can also be applied to processes of
permeation and pore opening in membranes (electroporation,
membrane disruption by antimicrobial peptides, vesicle fusion).
\end{abstract}
\date{\today}
\maketitle

\section{Introduction}
Many aspects of biological life crucially depend on stability of
cell membranes for which several properties are not understood
yet. Fluid lipid bilayers are the building blocks of biological
membranes. Pores in such systems play an important role in the
diffusion of small molecules across biomembranes \cite{PV96}. As
well pore formation is a possible mechanism for vesicle fusion
\cite{SK01}. In order for any vesicle to be useful it must be
relatively stable. Yet in order to undergo fusion, long-lived
holes must occur during the fusion transformation. How membranes
actually manage to exhibit these two conflicting properties is not
completely clear, but this can likely be realized only
dynamically. Dynamic properties are especially important for
biological membranes, because their static characteristics
describe a dead structure whereas life and biological functions
are associated with molecular motions. Thus, the desire to
understand the dynamics of biomembrane rupture, which is the main
aim of this paper, is hardly surprising.

On a microscopic level, pores are formed owing to thermal motion
of lipid molecules and, in principle, various types of pores can
be distinguished. It usually is assumed that initially nucleated
pores have hydrophobic edges; the so-called hydrophobic pores
\cite{GL88} which are spontaneously formed in the lipid matrix.
The probability for the existence of such hydrophobic pores is
determined by the free energy of pore as a function of pore radius
(see Sections \ref{prob} and \ref{anal} below). And, when the
hydrophobic pore exceeds a critical size, a reorientation of the
lipids takes place converting the pore into an hydrophilic one
where the head groups form the pore walls. As discussed in
\cite{TL03}, these reorientation processes can occur at the very
early stages of pore formation so that nucleation is the crucial
step in the rupture process. Note that thermal fluctuations can as
well lead to transient unstable pores, often termed as pre-pores
\cite{HU06}. In what follows, we will be interested on stable
pores (i. e., a well defined density of pores which can be
detected, e.g., by neutron scattering, with individual pores
forming and resealing reversibly) and head group reorientation
mechanisms will be disregarded. All details on hydrophobic and
hydrophilic pores will be lumped into the effective model
parameters like line tension, (unstressed) membrane surface
tension, and pore size diffusion coefficient. Electric breakdown
method provides information on pore size which can be drawn from
the dependence of membrane conductivity on applied voltage
\cite{GL88}, but dynamical pore characteristics can be hardly
found by this technique. One of the most relevant material
parameter controlling pore dynamics is membrane surface tension.
Surface tension suppressing thermal fluctuations and promoting
membrane adhesive properties, can induce adhesion of a membrane
onto a substrate or to another membrane, and other tension induced
morphological transitions including membrane rupture.

A closed vesicle without pores can survive for a very long period
of time. Pores can form and grow in the fluid-lipid membrane in
response to thermal fluctuations and external influences. Several
innovative techniques are available for observing transient
permeation and opening of pores. These include mechanical stress,
strong electric fields (electroporation), optical tweezers,
imploding bubbles, adhesion at a substrate, and puncturing by a
sharp tip. In all instances, the resulting transient pore is
usually unstable and leads to membrane rupture for some level of
the surface tension. Using the rupturing of biomembranes under
ramps of surface tension, the challenge of the Dynamic Tension
Spectroscopy (DTS) is to identify and quantify the relevant
parameters that govern the dynamics of membrane rupture and
thereby characterize the membrane mechanical strength.

\begin{figure}[ht]
\includegraphics[width=0.25\textwidth,angle=0]{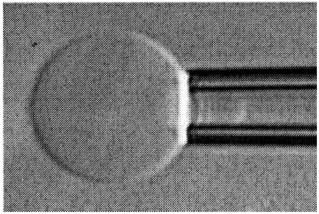}
\caption{Image of a 20  mm bilayer vesicle aspirated in a
micropipette (from Ref.~\cite{EHLR03}).} \label{fig1}
\end{figure}

As a demonstration of the DTS technique, Evans et
al.~\cite{EHLR03} conducted experiments of rupturing fluid
membrane vesicles with a steady ramp of micropipette suction
(Fig.~\ref{fig1}). Rupture tests on different types of giant
phosphatidychlorine lipid vesicles over loading rate
(tension/time) from 0.01 and up to 100 mN/m/s produce distributions of
breakage tensions governed by the kinetic process of membrane
failure. One might naively expect that lipid membranes to rupture
at tensions close to hydrocarbon - water surface tension as lipids
are held together by hydrophobic interactions. However,
biomembranes rupture at much lower tension. As pointed out by
Evans et al., rupture strength of a biomembrane is a dynamical
property and that the level of strength depends on the time frame
for breakage \cite{EHLR03}. Energy barriers along the tension
driven pathway are determinants of membrane strength, and the
relative heights of these barriers lead to time-dependent changes
in strength. To describe dynamic tension spectra they observed,
Evans et al. modeled the membrane breakage as a sequence of two
successive thermally-activated transitions (dependent of loading
rates) limited either by specific defect (pre-pore) formation or
by passage over the cavitation barrier (or evolution to an
unstable pore). Accordingly, this description was formalized into
a three-state kinetic model for the membrane: the defect-free
ground state, the defect or metastable state, and the ruptured
membrane state \cite{EHLR03}.

Motivated by these experimental and theoretical investigations and
findings, our objective in this paper is to develop a minimal
theoretical framework of the DTS method to describing the kinetic
process of membrane breakage. Based on the general framework of
Kramers reaction rate theory \cite{Kra40,HTB90}, we develop in
this paper a theoretical framework for DTS to describe the pore
growth and membrane rupture dynamics as a Markovian stochastic
process crossing a time-dependent energy barrier. As mentioned,
such a theoretical approach is conceptually similar to that used
by Evans et al., \cite{EHLR03} (see also, \cite{FJ03,BJ07}).
However, our description is more general than that presented in
\cite{EHLR03} as it characterizes and describes both primary
nucleation event followed by the continuous dynamics of pore
growth and shrinking allowing hence easy to follow and adapt for
further numerical treatments.

\section{Problem Formulation}
\label{prob}

In what follows, we treat the membrane as a two-dimensional
continuum medium and we neglect shape fluctuations, i.e., the
parameter $\delta \equiv k_B T/4 \pi \kappa $ \cite{SA94} is small
(e.g., $\delta\sim 10^{-3}$ for lipid bilayers). We will deal with
thermal fluctuations not related with shape fluctuations but with
the process of barrier crossing for pore formation characterized
by the parameter $\varepsilon= V_0/k_B T$, where $V_0$ is the
typical energy cost for pore formation at the critical pore radius
(see Eq.(\ref{pot}) below). Thus, our investigations will concern
the regime, $\delta\ll 10^{-1}\leq\varepsilon$.

Within the framework of the DTS, we describe the kinetic of
membrane rupture as a succession of two processes: an initial pore
nucleation followed by a diffusion dynamics of the pore size to
membrane rupture.

\subsection{Pore Formation}
For simplicity, we assume that the net process of pore nucleation
in a membrane can be described by an activated process following a
first-order kinetics with a rate $q$, i.e., the distribution of
times for the membrane to remain free of pores is given by the
exponential distribution with the rate, $q$, which is a function of
membrane characteristics. For the purpose of DTS, we will assume
that the pore nucleation rate $q(\sigma)$ is a function of
membrane surface tension $\sigma$.

\subsection{Pore Diffusion}
Once the pore is already formed in the membrane, the net energy
$V(r)$ of such a membrane of thickness $l$ with a circular pore of
radius $r$ consists of two opposed terms \cite{L75}: the surface
tension $\sigma$, favoring the expansion, and the energy cost
$\gamma$ of forming a pore edge (line tension), favoring the
closure:
\begin{equation}
V(r)=2\pi\gamma\,r-\pi\sigma\,r^2\:. \label{fener}
\end{equation}
Assuming that $\sigma>0$ and $\gamma>0$, and both are constant,
Eq.(\ref{fener}) predicts that for $r<a$, where $a=\gamma/\sigma$
is the pore radius for the maximum energy $V(r)$, the radial force
associated with a change in radius tends to reseal the pore, and
the membrane remains stable against pore growth. On the other
hand, a pore with a radius larger than the critical value $a$ will
grow without bound and, ultimately, will rupture the membrane. In
DTS experiments \cite{EHLR03}, the membrane is stressed such that
(provided that $\gamma$ remains constant) the surface tension
grows linearly with time as, $\sigma=\sigma_0+Ft$, where
$\sigma_0$ is the unstressed membrane tension and $F$ is the loading
rate constant. In this case, the critical radius $a(t)$ becomes a
decreasing function of time and any pore initially with radius
$r<a(0)$ will ultimately lead to membrane rupture at time such
that $r>a(t)$ as a result of the decreasing of both the critical
pore radius and associated barrier energy. Now, incorporating
thermal fluctuations in this picture, one can view the rupture of
the membrane as a Brownian process crossing the time-dependent
energy barrier $V[a(t)]$.

To setup the equations of pore size dynamics, we consider a
membrane with a pore of radius $r$ under mechanical stress that
changes its surface tension. In this description model of DTS, the
surface tension $\sigma$ in $V(r)$ is a linear function of time as
defined above. Thus, neglecting inertial effects, the dynamics of
$r$ is governed by the Langevin equation with time-dependent
potential,
\begin{eqnarray}
\left\{\begin{array}{l} \displaystyle
\zeta\,\frac{dr}{dt}=-\frac{dV(r,t)}{dr}+
f(t)\\\\
\displaystyle
V(r,t)=2\pi\gamma\,r-\pi\left(\sigma_0+Ft\right)\,r^2
\end{array}\right.
\label{lang1}
\end{eqnarray}
where $\zeta=4\pi\eta_m\,l$ is the friction coefficient to radial
circular fluctuations with $\eta_m$ the internal 2d membrane
viscosity, and $f(t)$ is a Gaussian random force of zero mean with
correlation function given by the fluctuation-dissipation
relation, $\langle f(t)f(t')\rangle=2\,\zeta\,{\rm
k_BT}\,\delta(t-t')$, with ${\rm k_BT}$ being the thermal energy.

\subsection{Dimensionless equations}

\begin{figure}[ht]
\includegraphics[width=0.475\textwidth,angle=0]{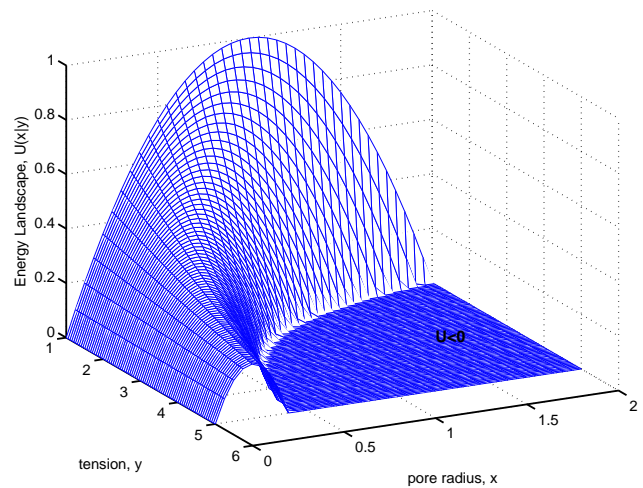}
\caption{Energy landscape (Eq.(\ref{pot}) with $\varepsilon=2$)
for the Dynamic Tension Spectroscopy. Flat area $x>2/y$
corresponds to $U<0$.} \label{fig2}
\end{figure}

To work with dimensionless variables, we define in Table~\ref{tb1}
scales of length, surface tension and time by $r_0$, $\sigma_0$
and $\tau$, respectively, and we consider the following
transformations: $x=r/r_0$, $y=\sigma/\sigma_0$, and $t\rightarrow
t/\tau$ (with $x\in [0,1]$ and $y\in [1,\infty[$) . This operation
leads us to define the control parameter,
\begin{equation}
v=\frac{F\tau}{\sigma_0}\equiv\frac{\mbox{diffusing time scale}}
{\mbox{surface tension time scale}}\:.
\end{equation}
This parameter allows us to distinguish two regimes in the
dynamics of the membrane rupture: the diffusion controlled regime
when $v\ll 1$ and the drift regime for $v\gg 1$ limit. With these
transformations Eq.(\ref{lang1}) can be rewritten as,
\begin{eqnarray}
\left\{\begin{array}{l}
\displaystyle\frac{dx}{dt}=-\frac{dU(x,y)}{dx}+X(t)\\ \\
\displaystyle\frac{dy}{dt}=v
\end{array}\right.
\label{lang2}
\end{eqnarray}
where $X(t)$ is a Gaussian random force of zero mean with
correlation function given by, $\langle
X(t)X(t')\rangle=2\,\delta(t-t')$, and we have defined the
potential (energy landscape for DTS illustrated in
Fig~\ref{fig2}),
\begin{equation}
U(x|y)=\frac{\varepsilon}{2}\,\left[2x-yx^2\right]\:\:;\:\:\varepsilon=\frac{V(r_0)}{\rm
k_BT}=\frac{\pi\gamma^2}{\sigma_0{\rm k_BT}}\:.\label{pot}
\end{equation}
The potential $U(x|y)$ is maximum at $x^{\ddag}=1/y$ corresponding
to the energy barrier $U^{\ddag}=U(x^{\ddag}|y)=\varepsilon/2y$.
Both the position $x^{\ddag}$ and height $U^{\ddag}$ of the energy
barrier decrease as $y$ gets larger as a result of the membrane
stress.

\begingroup \squeezetable
\begin{table}
\caption{Parameters and dimensionless variables. The index "0"
denotes quantities for unstressed membrane.}
\begin{tabular}{ll}  \hline\hline
Symbols & Definition \\ \hline
$\gamma$ & line tension ({\it energy/length}) \\
$\sigma_0$ & unstressed surface tension ({\it energy/surface}) \\
$F$ & tension loading rate ({\it energy/surface/time}) \\
$D={\rm k_BT}/\zeta\,l$ & pore diffusion coefficient ({\it
length$^2$/time})
\\\hline
$r_0=\gamma/\sigma_0$ & critical pore radius ({\it length}) \\
$\tau=r_0^2/D$ & diffusing time scale of the critical pore ({\it time}) \\
$F_0=\sigma_0/\tau$ & critical tension loading rate ({\it
energy/surface/time})
\\\hline
$q_0$ & reduced unstressed pore nucleation rate \\
$x=r/r_0$ & reduced pore radius \\
$y=\sigma/\sigma_0$ & reduced membrane surface tension \\
$\varepsilon=\pi\gamma^2/\sigma_0{\rm k_BT}$ & reduced energy barrier for unstressed membrane \\
$v=F/F_0$ & reduced tension loading rate:$\left\{\begin{array}{l}
\!v<1\,:\,\mbox{diffusing limit}\\
\!v>1\,:\,\mbox{drift limit}
\end{array}\right.$\\
\hline \hline
\end{tabular}
\vspace{0.5cm} \label{tb1}
\end{table}
\endgroup

\subsection{DTS observables}
As the barrier crossing to both pore nucleation and membrane
rupture are stochastic processes, both the membrane lifetime and
the membrane tension at rupture are distributed. Our goal is to
calculate the two quantities that characterize the kinetics of
membrane rupture in DTS framework: the rate of membrane rupture
and the distribution of tension at membrane rupture.

For a membrane with a pore in the absence of mechanical stress
$v=0$, the distribution of tension at membrane rupture is delta
function, $Q(y)=\delta(y-1)$, and the rate of membrane rupture can
be obtained analytically using the first passage time
approach~\cite{SSS80,BS97},
\begin{eqnarray}
& & \frac{1}{k(\varepsilon|0)}=\int_{0}^{1}\!\frac{dx}{p_{\rm
eq}(x|1)}\,
\left[\int_{0}^{x}\!p_{\rm eq}(z|1)\,dz\right]^2 \nonumber \\
& = & \frac{\sqrt{\pi/(2\varepsilon)}} {{\rm
erfi}\left[\sqrt{\varepsilon/2}\right]} \int_{0}^{1}\!dx\,{\rm
e}^{-\varepsilon (x-1)^2/(2)}
\nonumber\\
& & \times\left\{{\rm erfi}\left[\sqrt{\varepsilon/2}\right]- {\rm
erfi}\left[(1-x)\sqrt{\varepsilon/2}\right]\right\}^2 \label{mfpt}
\end{eqnarray}
in which we have assumed that the membrane system was initially
prepared with the distribution $p_{\rm eq}(x|1)$, where
\begin{eqnarray}\label{peq}
p_{\rm eq}(x|y)=\frac{{\rm
e}^{-U(x|y)}}{Z(y)}\:;\:Z(y)=\int_{0}^{1/y}\!{\rm
e}^{-U(x|y)}\,dx\:,
\end{eqnarray}
and $Z(y)=\displaystyle\left(\frac{\pi {\rm
e}^{-\varepsilon/y}}{2\varepsilon y}\right)^{1/2}\,{\rm
erfi}\left[\sqrt{\varepsilon/2y}\right]$ where ${\rm erfi}[z]={\rm
erf}[iz]/i$ and ${\rm erf}[\cdots]$ is the error function.

On the other hand, for a membrane initially free of pore and for
$v>0$, analytical expressions are not straightforward but the rate
of membrane rupture and the distribution of tension at membrane
rupture can be determined as follows. Let $S(t)$ be the survival
probability that describes the fate of the membrane from the
beginning of the experiment. The distribution of membrane lifetime
or rupture time is given by $-dS/dt$, and the membrane rupture
rate (equals to the inverse of the membrane lifetime) is obtained
as,
\begin{equation}\label{tau}
\frac{1}{k(\varepsilon|v)}=\int_{0}^{\infty}\!t\left[-\frac{dS(t)}{dt}\right]\,dt=\int_{0}^{\infty}\!S(t)\,dt\:.
\end{equation}
Likewise, the distribution $Q(y)$ of tensions $y$ at which the
membrane rupture is related to the distribution of rupture time
and, as $y=1+vt$, we have:
\begin{equation}\label{Qofy}
Q(y)=\left.\left|\frac{dt}{dy}\right|\times\left(-\frac{dS}{dt}\right)\right|_{y=1+vt}.
\end{equation}
The DTS spectrum (here, mean of rupture tensions), $\langle
y(\varepsilon|v)\rangle=\displaystyle\int_{1}^{\infty}\!yQ(y)dy$,
is related to the rupture rate $k(\varepsilon|v)$ by, $\langle
y(\varepsilon|v)\rangle=1+v/k(\varepsilon|v)$. In the case where
$S(t)$ satisfies a first-order rate equation with the
time-dependent rate $\Gamma(t)$, i. e.,
$S(t)=\exp\left\{-\displaystyle\int_{0}^{t}\!\Gamma(t')dt'\right\}$,
the distribution $Q(y)$ can written as,
\begin{eqnarray}
Q(y)=\frac{\Gamma\left(\frac{y-1}{v}\right)}{v}\,\exp\left\{-\int_{0}^{(y-1)/v}\!\Gamma(z)\,dz\right\}\:.
\end{eqnarray}
It follows from this that $S(t)$ is the key function to derivation
of expressions of quantities of interest.

\section{Analytical Theory}
\label{anal}

Equivalently to the stochastic equation in Eq.(\ref{lang2}), the
joint probability density, $P(x,y,t)$, of finding the membrane
with surface tension $y$ and pore of radius $x$ (i.e., the phase
space point $(x,y)$) at time $t$ is described by the
two-dimensional Fokker-Planck equation:
\begin{equation}
\frac{\partial P(x,y,t)}{\partial t}=-v\, \frac{\partial
P(x,y,t)}{\partial y}\,-\, \frac{\partial J(x,y,t)}{\partial x}\:,
\label{FP}
\end{equation}
where the first term in the right hand side describes the
ballistic drift of the surface tension caused by the applied
loading rate, and the second term is the diffusive flux describing
the diffusion of the pore radius in the potential $U(x|y)$ for
given $y$,
\begin{equation}
J(x,y,t)=-\,{\rm e}^{-U(x|y)}\, \frac{\partial}{\partial x}\,{\rm
e}^{U(x|y)}\,P(x,y,t)\:. \label{flux}
\end{equation}
Eq.(\ref{FP}) reduces to the Smoluchowski equation in the $v=0$
limit \cite{ZW01}. To study the rupture of membrane as an escape
of the pore radius undergoing a Brownian dynamics within the
interval $[0,1[$, we require that $P(x,y,t)$ satisfies the
reflecting boundary condition at $x=0$ and the absorbing boundary
condition at $x=1/y$, i.e.:
\begin{eqnarray}
\left\{\begin{array}{lcl}
J(x,y,t)=0 & \mbox{at} & x=0\:,\\
P(x,y,t)=0 & \mbox{at} & x=1/y\:.
\end{array}\right.
\label{bc}
\end{eqnarray}
Formal, yet numerically computable, solution of Eq.(\ref{FP}) with
the initial condition
$P(x,y,t=t_0|x_0,y_0)=\delta(x-x_0)\,\delta(y-y_0)$ and boundary
conditions in Eq.(\ref{bc}) is given by the Green's function,
\begin{eqnarray}
& & P(x,y,t|x_0,y_0,t_0)=\left[\frac{p_{\rm eq}(x|y)}{p_{\rm
eq}(x_0|y_0)}\right]^{1/2}\,\delta\left[y-y_0-v(t-t_0)\right]\nonumber \\
& & \times \sum_{n=1}^{\infty}\psi_n(x_0|y_0)\,\psi_n(x|y)\,
\exp\left\{-\frac{1}{v}\int_{y_0}^{y}\!\lambda_n(z)\,dz\right\}.
\label{Green}
\end{eqnarray}
The $\psi_n(x|y)$ and $\lambda_n(y)$ are respectively the
normalized eigenfunctions $\left(\displaystyle
\int_{0}^{1/y}\!dx\,\psi_n(x)\psi_{n'}(x)=\delta_{n,n'}\right)$
and associated eigenvalues of the eigenvalue problem,
\begin{eqnarray}
H\psi=\frac{d^2\psi}{dx^2}-\left[\frac{v\varepsilon
x^2}{4}+\frac{\varepsilon^2(1-y x)^2}{4}+\frac{\varepsilon
y}{2}\right]\psi=-\lambda\psi\:,
\end{eqnarray}
satisfying the reflecting and absorbing boundary conditions at
$x=0$ and $x=1/y$, respectively,
\begin{eqnarray}\label{bc2}
\left\{\begin{array}{l} \displaystyle\left.{\rm
e}^{-U(x)}\,\frac{\partial}{\partial x}\left[\,{\rm e}^{U(x)/2}\,\psi(x)\right]\right|_{x=0}=0\\
 \\
\psi(x=1/y)=0
\end{array}\right.
\end{eqnarray}
Let $z=\displaystyle\left[v\varepsilon+\varepsilon^2
y^2\right]^{1/4}\,\left[x-\frac{\varepsilon y}{v+\varepsilon
y^2}\right]$ such that $z_0\leq z\leq z_1$, where,
\begin{equation}\label{z01}
z_0=-\frac{\varepsilon y\left[v\varepsilon+\varepsilon^2
y^2\right]^{1/4}}{(v+\varepsilon
y^2)}\:\:;\:\:z_1=\frac{v\left[v\varepsilon+\varepsilon^2
y^2\right]^{1/4}}{y(v+\varepsilon y^2)}\:.
\end{equation}
The eigenvalue problem, $H\psi=-\lambda\psi$, becomes,
\begin{eqnarray}\label{ham}
\left\{\begin{array}{l}
\displaystyle\frac{d^2\psi}{dx^2}+\left[E-\frac{z^2}{4}\right]\psi=0\\
\displaystyle E=\frac{1}{\left[v\varepsilon+\varepsilon^2
y^2\right]^{1/2}}\,\left[\lambda-\left(\frac{\varepsilon
y}{2}+\frac{v\varepsilon^2}{4(v+\varepsilon y^2)}\right)\right]\
\end{array}\right.
\end{eqnarray}
The general solution to Eq.(\ref{ham}) which satisfies the
absorbing boundary condition in Eq.(\ref{bc2}) is given by,
\begin{eqnarray}
\left\{\begin{array}{l}
\displaystyle \psi(z)=A\,\left[D_{\nu}(-z_1)D_{\nu}(z)-D_{\nu}(z_1)D_{\nu}(-z)\right]\\
\displaystyle \nu=E-\frac{1}{2}\:
\end{array}\right.
\end{eqnarray}
where $D_{\nu}(z)$ is the Weber's parabolic cylinder function
\cite{AS72}. The constant $A$ is obtained from the normalization
and the eigenvalue spectrum by solving the following eigenvalue
equation obtained be using the reflecting boundary in
Eq.(\ref{bc2}),
\begin{eqnarray}\label{eveq}
& &
D_{\nu}(-z_1)\left[\frac{dD_{\nu}(z_0)}{dz_0}+\frac{\varepsilon}{2\left[v\varepsilon+\varepsilon^2
y^2\right]^{1/4}}\,
D_{\nu}(z_0)\right]+\nonumber\\
& &
D_{\nu}(z_1)\left[\frac{dD_{\nu}(-z_0)}{dz_0}-\frac{\varepsilon}{2\left[v\varepsilon+\varepsilon^2
y^2\right]^{1/4}}\, D_{\nu}(-z_0)\right]\nonumber\\
& & =0\:.
\end{eqnarray}

Now, assuming that the system is initially prepared with the
distribution $g(x,y,t)$ describing the pore formation, the
survival probability that describes the fate of the membrane with
a pore is given by,
\begin{eqnarray}
S(t) & = &
\int_{1}^{\infty}\!dy_0\int_{1}^{\infty}\!dy\int_{0}^{1/y_0}\!dx_0\int_{0}^{1/y}\!dx\,\int_{0}^{t}\!dt_0\nonumber\\
& & P(x,y,t|x_0,y_0,t_0)\,g(x_0,y_0,t_0)\:, \label{sur}
\end{eqnarray}
where $P(x,y,t)$ is given in Eq.(\ref{Green}) and the preparation
distribution,
\begin{eqnarray}\label{prep}
g(x,y,t) & = & p_{\rm eq}(x|y)\,\delta(y-1-vt)\nonumber\\
& & \times\left[q(t)\exp\left\{-\int_{0}^{t}\!q(t')dt'
\right\}\right]\:,
\end{eqnarray}
where the term between squared  brackets stands for the
distribution of times for pore nucleation.

Equation (\ref{sur}) provides an exact expression of $S(t)$ in
terms of infinite series from which the rupture rate
$k(\varepsilon|v)$ (or the DTS spectrum) and the distribution
$Q(y)$ of rupture tension can subsequently derived by using
Eqs.(\ref{tau}) and (\ref{Qofy}), respectively, and approximate
expression for $\Gamma(t)$ as well.

Interestingly, these derivations can also be used to establish the
correspondence with the three-state model in Ref.~\cite{EHLR03}
and, therefore, provide exact expressions as,
\begin{eqnarray}\label{3states}
\left\{\begin{array}{lcl}
S_{\circ}(t) & = &\int_{1}^{\infty}\!dy_0\int_{1}^{\infty}\!dy\int_{0}^{1/y_0}\!dx_0\int_{0}^{t}\!dt_0\\
& & P(0,y,t|x_0,y_0,t_0)\,g(x_0,y_0,t_0)\\
S_{\ast}(t) & = &S(t) - S_{\circ}(t)\\
S_{\rm hole}(t) & = &1 - S(t)
\end{array}
\right.
\end{eqnarray}
where $S_{\circ}(t)$ is the defect-free ground state,
$S_{\ast}(t)$ the defect or metastable state, and  $S_{\rm
hole}(t)$ the ruptured membrane state as defined in
Ref.~\cite{EHLR03}.

Unfortunately, derivation of analytical expressions (which require
solving Eq.(\ref{eveq})) as outlined above may be tedious and
obtained results turn out not easy to use in practice. These
calculations were done mainly for the purpose of presenting the
derivation formalism of exact expressions. Such exact solutions
may turn out useful for checking simulation results just like
those presented in the next section. Our main interest in this
paper is to understand, write down equations describing DTS
experiments and develop related simulations that could be compared
with experimental data. To this end, we switch to the simulations
of the kinetics of the membrane rupture as described by stochastic
and dynamical equations outlined above.

\section{Simulation Algorithm}

The simulations of the kinetics of the membrane rupture were
performed using the discretized version of Eq.(\ref{lang2}) to
have the algorithm,
\begin{eqnarray}
\left\{\begin{array}{l}
x_{n+1}=(\varepsilon\Delta y_n+1)x_n-\varepsilon\Delta+X_n\\
y_{n+1}=y_n+v\Delta
\end{array}\right.
\label{langsim}
\end{eqnarray}
where $\Delta$ is the time step and the Gaussian random noise
$X_n$ is defined by the moments, $\langle X_n\rangle=0$ and
$\langle X_nX_{n'}\rangle =2\Delta\,\delta_{nn'}$. For each
trajectory for a membrane free of pore at $t=0$, with fixed
barrier height $\varepsilon$ and loading rate $v$, the system is
prepared according to the distribution $g(x_0,y_0,t_0)$ given in
Eq.(\ref{prep}): the initial pore is created in the membrane at
time $t_0=(y_0-1)/v$ where the membrane surface tension $y_0$ for
pore creation is given by the exponential distribution,
\begin{equation}\label{fofy}
f(y)=\left(\frac{q(y)}{v}\right)\exp\left\{-\int_{1}^{y}\!\left(\frac{q(y')}{v}\right)dy'
\right\}\:,
\end{equation}
and the pore radius $x_0$ is generated from the distribution
$p_{\rm eq}(x_0|y_0)$ in Eq.(\ref{peq}). From this, the next pore
radii $x_n$ and surface tensions $y_n$ are generated according to
the algorithm in Eq.(\ref{langsim}). To simulate the rupture of
the membrane, each trajectory starting at $x_0$ ($0\leq
x_0<1/y_0$) at time $t=t_0$ is terminated at time $t_i=n \Delta$
when the condition $x_n\geq 1/y_n$ is satisfied for the first time
(the boundary at $x=0$ is reflecting). The rupture surface tension
$y_i=y_n$, the first passage time $t_i$ and the survival
probability $S_i(t)$ (defined as $S_i(t)=1$ for all $t<t_i$ and
$S_i(t)=0$ otherwise) for this given trajectory are recorded. The
distribution $Q(y)$ of rupture tensions is obtained by binning the
$y_i$'s over a large number $N$ of trajectories. Likewise, the
definitive survival probability, $S(t)$, and the rupture rate
constant, $k(\varepsilon|v)$, (i.e., the inverse of the membrane
mean lifetime) are then obtained by averaging these quantities
over a large number of trajectories:
\begin{equation}
S(t)=\frac{1}{N}\sum_{i=1}^{N}S_i(t)\:\:\:\:\mbox{and}\:\:\:\:
\frac{1}{k(\varepsilon|v)}=\frac{1}{N}\sum_{i=1}^{N}t_i\:.
\end{equation}
For all simulations reported in this paper we used
$\Delta=10^{-5}$ and a total of $N=10^5$ trajectories were used to
perform the averages.

\section{Results}

In what follows, simulations were carried out with the
tension-dependent pore nucleation rate given by, $\displaystyle
q(y)=q_0{\rm e}^{\alpha(y-1)}$, where $q_0$ is the pore creation
rate in the unstressed membrane and $\alpha$ is a constant
depending on membrane characteristics and temperature (e.g.,
$\varepsilon$). As the membrane tension increases with time with
load $v$ the overall rate of pore formation will be given by,
\begin{equation}\label{nrt}
k_n=q_0\,\left[\int_{0}^{\infty}\!dx\,\exp\left\{-\frac{\left({\rm
e}^{at}-1\right)}{a}\right\}\right]^{-1}\:;\:a=\frac{\alpha
v}{q_0}\:.
\end{equation}
The DTS outputs are the distribution $Q(y)$ of tensions at rupture
and the DTS spectrum defined by the plot of the mode of $Q(y)$ as
a function of $ln(v)$ \cite{EHLR03}. We have computed the membrane
survival probability (results not reported) distribution $Q(y)$
and rupture rate $k(\varepsilon|v)$. As a successive process, the
membrane rupture rate can be written as, $k(\varepsilon|v)=k_d
k_{\rm n,eff}/\left(k_d+k_{\rm n,eff}\right)$ where $k_{\rm
n,eff}$ (different from $k_n$) is the effective rate of pore
formation and $k_d$ is the rupture rate for a membrane initially
with a pore in it. In what follows, we will investigate the effect
of $q$, $v$ and $\varepsilon$ and on $Q(y)$ and
$k(\varepsilon|v)$.

\subsection{Diffusion Controlled Limit: $q\rightarrow\infty$ limit}

\begin{figure}[ht]
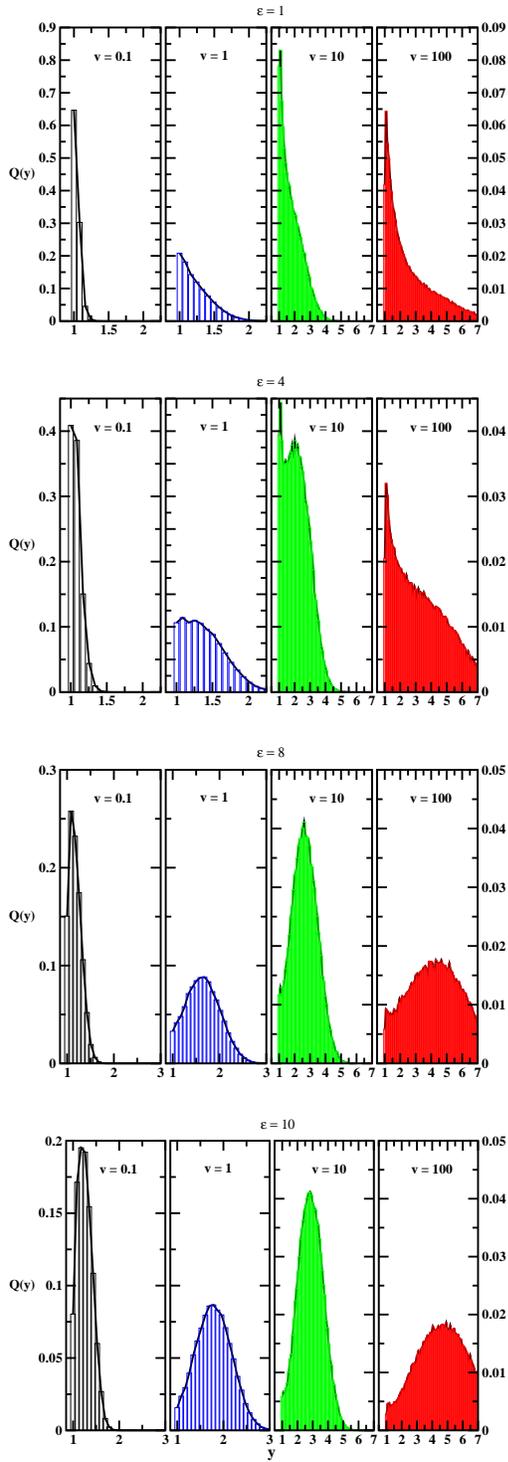

\vbox{\includegraphics[width=0.37\textwidth,angle=0]{fig3a.eps}\vspace{0.27cm}
\includegraphics[width=0.37\textwidth,angle=0]{fig3b.eps}\vspace{0.27cm}
\includegraphics[width=0.37\textwidth,angle=0]{fig3c.eps}\vspace{0.27cm}
\includegraphics[width=0.37\textwidth,angle=0]{fig3d.eps}}
\caption{Distribution, $Q(y)$, of rupture tensions, $y$, for
various values of the energy barrier, $\varepsilon$, and loading
rate, $v$. For each row of figures, the leftmost and second
leftmost figures have the same scales in both $x$ and $y$ axes,
and likewise for rightmost and second rightmost figures. Note that
$y$ - scales of rightmost figures are an order of magnitude
smaller than for the leftmost ones.} \label{fig3}
\end{figure}

\begin{figure}[ht]
\includegraphics[width=0.37\textwidth,angle=0]{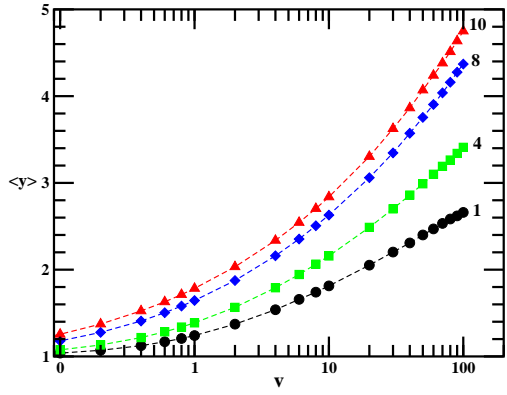}
\caption{DTS spectrum $\langle y(\varepsilon|v)\rangle$ as a
function of the loading rates $v$ (in logarithm scale) for various
barrier heights $\varepsilon$ (quoted numbers). Filled symbols
correspond to simulations and dashed lines are guide eyes.}
\label{fig4}
\end{figure}

\begin{figure}[ht]
\includegraphics[width=0.37\textwidth,angle=0]{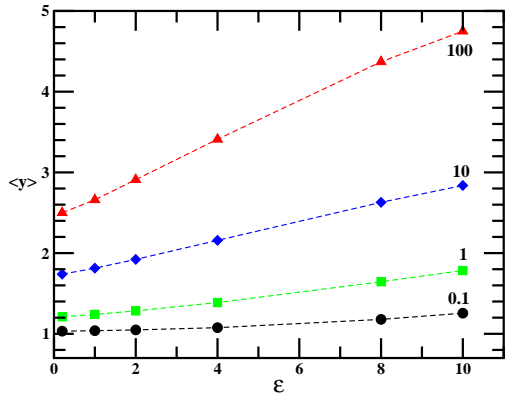}
\caption{DTS spectrum $\langle y(\varepsilon|v)\rangle$ as a
function of the barrier height $\varepsilon$ for various loading
rates $v$ (quoted numbers). Filled symbols correspond to
simulations and dashed lines are guide eyes.} \label{fig5}
\end{figure}

This limit corresponds to the situation where membrane under
tension stress has already a pore in it and, therefore,
$k(\varepsilon|v)=k_d$. Figures~\ref{fig3}, \ref{fig4} and
\ref{fig5} display the distribution $Q(y)$ and the DTS spectrum as
a function of the energy barrier $\varepsilon$ and loading tension
rate $v$. As can be seen, the distribution $Q(y)$ of tensions at
rupture broaden from the delta function at $y=1$ to a wider
distribution when both $v$ and $\varepsilon$ increase.
Accordingly, the mean tensions $\langle y(\varepsilon|v)\rangle$
for membrane rupture increase with both loading rate $v$ and
barrier height $\varepsilon$.

\subsection{Finite Pore Nucleation Rate $q\neq 0$ limit}

To investigate the effect of pore nucleation rate on the membrane
rupture, we consider two cases of increasing complexity.

\subsubsection{$\alpha=0$ limit}

\begin{figure}[ht]
\includegraphics[width=0.37\textwidth,angle=0]{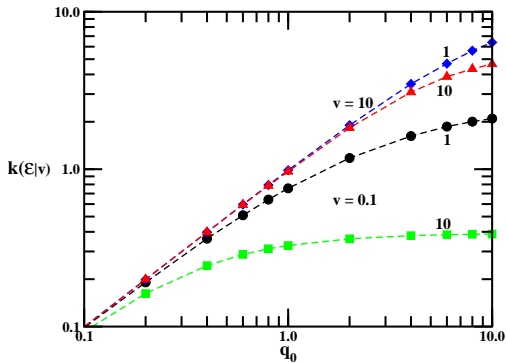}
\caption{Rupture rate $k(\varepsilon|v)$ as a function of the
tension independent pore nucleation rate $q_0$ (for $\alpha=0$)
for various barrier heights $\varepsilon$ (quoted numbers) and
loading rates $v=0.1$ (circles and squares) and $v=10$ (diamonds
and triangles). Filled symbols correspond to simulations and
dashed lines are guide eyes.} \label{fig6}
\end{figure}

\begin{figure}[ht]
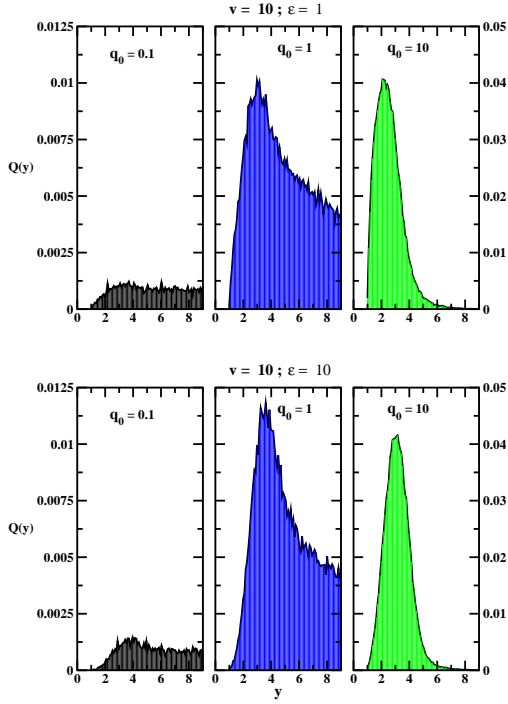

\vbox{\includegraphics[width=0.37\textwidth,angle=0]{fig7a.eps}\vspace{0.27cm}
\includegraphics[width=0.37\textwidth,angle=0]{fig7b.eps}}
\caption{Distribution, $Q(y)$, of rupture tensions, $y$, for
various values of the tension independent pore nucleation rate,
$q_0$ (for $\alpha=0$), and loading rate, $v$. For each row of
figures, panels with $q_0=1$ and $q_0=10$ have the same scales in
both $x$ and $y$ axes.} \label{fig7}
\end{figure}

In this case, $k_{\rm n,eff}=k_{\rm n}=q_0$. Figure~\ref{fig6}
shows the variation of rupture rate $k(\varepsilon|v)$ as a
function of the pore nucleation rate $q_0$. As expected for a
successive process, $k(\varepsilon|v)$ linearly grows with $q_0$
in the nucleation controlled limit when $q_0\ll k_d$ and saturates
to $k_d$ in the diffusion controlled limit for $q_0\gg k_d$.
Figure~\ref{fig6} illustrates that system parameters can be tuned
to follow the transition between the nucleation and diffusion
controlled limits. Accordingly, Fig.~\ref{fig7} displays the
profiles of the distribution $Q(y)$ corresponding to the
nucleation controlled limit and toward the diffusion controlled
limit.

\subsubsection{Case of $\alpha=1$}

\begin{figure}[ht]
\includegraphics[width=0.37\textwidth,angle=0]{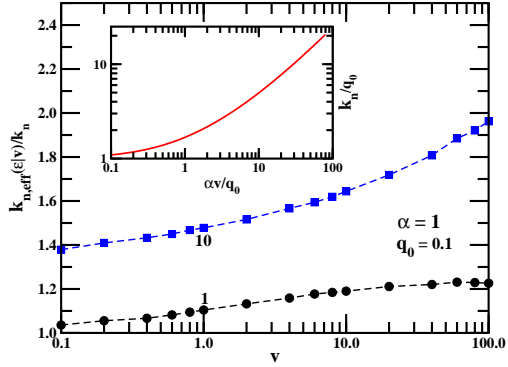}
\caption{Reduced effective nucleation rate $k_{\rm
n,eff}(\varepsilon|v)/k_n(v)$ as a function of the loading rates
$v$ in log-linear scale for various barrier heights $\varepsilon$
(quoted numbers), with $q_0=0.1$ and $\alpha=1$. Filled symbols
correspond to simulations and dashed lines are guide eyes.
\textbf{Inset:} Reduced nucleation rate $k_n/q_0$ versus the
variable $\alpha v/q_0$ in log-log scale.} \label{fig8}
\end{figure}

\begin{figure}[ht]
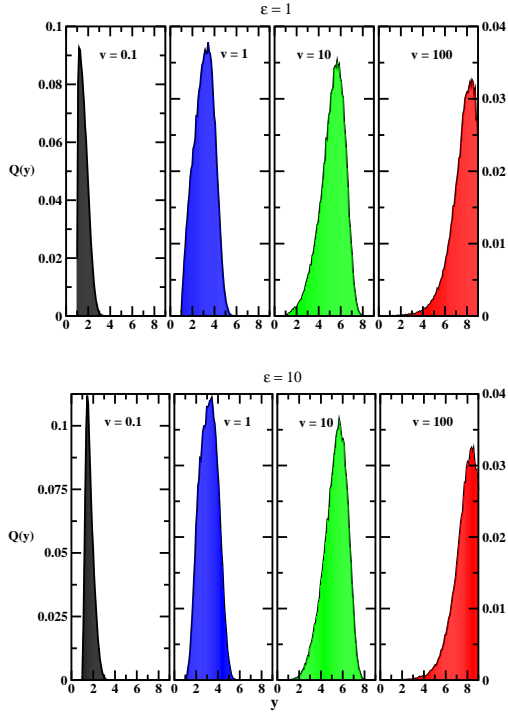

\vbox{\includegraphics[width=0.37\textwidth,angle=0]{fig9a.eps}\vspace{0.27cm}
\includegraphics[width=0.37\textwidth,angle=0]{fig9b.eps}}
\caption{Distribution, $Q(y)$, of rupture tensions, $y$, for
various values of loading rate, $v$, and energy barrier,
$\varepsilon$, with $q_0=0.1$ and $\alpha=1$. For each row of
figures, panels with $v=1$, $v=10$ and $v=100$ have the same
scales in both $x$ and $y$ axes.} \label{fig9}
\end{figure}

When $\alpha\neq 0$, the dynamic nature of membrane rupture leads
to $k_{\rm n,eff}\neq k_{\rm n}$. Both $k(\varepsilon|v)$ and
$Q(y)$ still exhibit similar behaviors observed in the case of
$\alpha=0$ but with nontrivial dependence on the loading rate $v$
and barrier height $\varepsilon$. As $k_d(\varepsilon|v)$ is known
from the limit $q\rightarrow\infty$, the behavior of $k_{\rm
n,eff}(\varepsilon|v)$ can be learned from Fig.~\ref{fig8}.
Clearly, $k_{\rm n,eff}\geq k_{\rm n}$, and the effective pore
nucleation rate has extra $v$ and $\varepsilon$ dependencies that
are not taken into account in $k_{\rm n}$ in the absence of pore
dynamics. The departure of $k_{\rm n,eff}$ from $k_{\rm n}$
increases with both $v$ and $\varepsilon$ indicating that opening
pore is more likely for high barrier membrane with high loading
rates. The kind of distributions $Q(y)$ that can be observed in
this limit are displayed in Fig.~\ref{fig9}.

\section{Concluding Remarks}

Rupturing fluid membranes or vesicles with a steady ramp of
micropipette suction produces a distribution of breakage tensions
governed by the kinetic process. Experimental evidences have
demonstrated that the membrane rupture is a dynamical property
whose the strength depends on the time scale for breakage. From
the theoretical point of view, we have developed a minimal model
for the Dynamic Tension Spectroscopy that describes the pore
nucleation as a first-order activated process, the dynamics of
pore growth as a two-dimensional (in pore radius and tension
spaces) Markovian stochastic process, and the rupturing of
membrane is modeled by an escape process over the time-dependent
critical barrier of the energy landscape. We have provided an
exact analytical solution of this problem and established the
correspondence between this description and the three-state model
in Ref.~\cite{EHLR03}. As numerical results, we have simulated the
rupture rate and the distribution of rupture tension as a function
of the pore nucleation rate, the critical barrier height,
$\varepsilon$, of the unstressed membrane and the reduced tension
loading rate, $v$. Our simulated histograms reproduce already
several features observed in DTS experiments in \cite{EHLR03} and
highlight the variety of profiles and richness of the problem.
Indeed, the distribution of rupture tensions show different
profiles between and in the two nucleation and diffusion
controlled limits as a function of $\varepsilon$ and $v$.

As presented above, the kinetic of membrane rupture as probed in
DTS experiments is very similar to non-equilibrium problems
studied in single-molecule pulling experiments using atomic-force
microscopes \cite{BQG86,MA88}. To cite a few, there are several
theoretical works \cite{BE78,ER97,ISB97,HS03,OHS06} that have been
developed, extended and refined to describe the thermally
activated rupture events within the general framework of Kramers
reaction rate theory \cite{Kra40,HTB90}.

Needless to recall that the theoretical model outlined above does
not yet take into account all aspects of the membrane rupture
observable in the experiments since we purposely neglected some of
features like, e.g., the coarse-grained structure of the membrane.
However, the developed formulation can be embellished in several
directions to include the mentioned above and some other
ingredients, like, non-Markovian dynamics of the pore radius
dynamics driven by the membrane matrix in which the pore is
embedded, and eventually, the time variation of barrier energy
$\varepsilon$ due to change of the line tension. Such a
generalized reaction rate approach can be also applied to membrane
disruption by antimicrobial peptides. Indeed a peptide binding
causes a local membrane area expansion and therefore it is
equivalent to local tension (see more about the problem in
\cite{HU06,BJ07}). Such a work is underway.

\end{document}